\documentclass[a4wide,12pt]{article}
\usepackage[margin=1.2in]{geometry}
\usepackage[utf8]{inputenc}
\usepackage [english] {babel} 
\usepackage{amsmath,amssymb,latexsym}
\usepackage{slashed}
\usepackage{graphicx}
\usepackage{epstopdf}
\usepackage{relsize}
\usepackage{mathtools}
\usepackage{ytableau}
\usepackage{subcaption}
\usepackage[sorting=none, backend=bibtex]{biblatex}
\bibliography{ref}
\DeclareGraphicsExtensions{.png}

\usepackage{lmodern}

\usepackage{hyperref}
\usepackage{pict2e}

\def \sec{\begin{section}}
\def \esec{\end{section}}

\def \la {\lambda}

\def \om {\omega}

\def \pr {\partial}
\def \ra {\rightarrow}

\def \mc {\mathcal{M}}
\def \cc {\mathcal{C}}
\def \dc {\mathcal{D}}
\def \lc {\mathcal{L}}

\def \bb {\mathbb{B}}

\def \beq { \begin{equation}}
\def \eeq {\end{equation}}

\renewcommand\Re{\operatorname{Re}}

\def \l {\left(}
\def \r {\right)}

\def \inst {\text{inst}}

\begin{document}

\vskip 2cm

\begin{center}
{\Large \bf
On Lagrangian description of Borel resummation
}

\vskip 1.0cm

A. Milekhin\footnotetext{email: milekhin@princeton.edu}
\bigskip
\bigskip

\begin{tabular}{c}
Department of Physics, Princeton University, 08540, Princeton, NJ, USA\\
Institute for Information Transmission Problems of Russian Academy of Science, \\
B.~Karetnyi 19, Moscow 127051, Russia\\
Institute of Theoretical and Experimental Physics, B.~Cheryomushkinskaya 25,\\
Moscow 117218, Russia
\end{tabular}

\vskip 1.0cm

\textbf{Abstract}
\end{center}

\medskip
\noindent
In this paper we will propose a physical interpretation of Borel resummation.
We will define Borel transformation on the Lagrangian level. This 
transformation essentially coincides with coupling the original theory to the non-dynamical 
complex dilaton field. In the perturbative sector it reproduces standard Borel resummation. In the non-perturbative sector 
this transformation reproduces Bogomolny--Zinn-Justin(BZJ) regularization. Our approach unifies
the perturbative Borel resummation and non-perturbative BZJ procedure.

\bigskip

\section{Introduction}
The main motivation of this paper is to answer the following question: what is the physical picture behind 
the Borel resummation of perturbation expansion? We will propose the following answer: it is equivalent to coupling
the original theory to a non-dynamical complex scalar field. This field strikingly resembles 
a dilaton field. However the integration contour for this field is non-standard and purely complex.  
Non-standard integration cycles in path integrals
have been intensively discussed in recent years\cite{wqm,wcs,jp,pf,du} and it is not something new. 

An interesting feature 
arises when we study the non-perturbative sector of the transformed theory. 
In the (multi-)instanton background one has to
carefully choose the integration cycle for the complex dilaton field. To illustrate it, we studied a simple quantum 
mechanical problem of particle in a double-well potential. As it turns out the integration cycle for 
the instanton--anti-instanton
background is unique and reproduces the so-called Bogomolny--Zinn-Justin regularization. We will explain in a minute what
BZJ prescription is, but for now, let us mention that it 
includes an analytical continuation of the coupling constant. In our approach
the coupling constant will stay real, but the integration cycle for the dilaton field will be complex.

We expect that the non-perturbative sector of the transformed theory is richer than in the original theory.
We will illustrate this using a simple field-theoretic model of a massless scalar field with self-interaction. 
Let us also mention that our approach allows to systematically generalize the BZJ prescription for the case of several coupling 
constants. We will explain it in more detail in the next section.

Let us now describe in more detail what is 
Borel regularization and what me mean by the BZJ prescription.

Suppose we want to calculate an amplitude $\mc$ in some quantum mechanical problem or quantum field theory described by the following partition function:
\beq
Z = \int \ \dc x \exp(i S_0(x) +i S_i(x,g_0))
\eeq
where $x$ denotes the set of fields, $S_0$ is a free action and $S_i$ is interaction term which depends on the coupling constant $g_0$.
Perturbation theory gives the answer as a Taylor series in the coupling constant:
\begin{equation}
\label{mc}
\mc=\sum_{n=0}^{+\infty} c_n g_0^n
\end{equation}
Unfortunately, in most cases this series has zero radius of convergence: typical behavior  of $c_n$ is $c_n \sim n! A^n$, 
where $A$ is some constant.
One can try to make sense out of this expansion by using the so-called Borel resummation. Instead of of the original series one considers Borel-regularized series:
\begin{equation}
\label{mcreg}
\mc^\bb(g) = \sum_{n=0}^{+\infty} \dfrac{c_n g^n}{n!}
\end{equation}
where we have introduced Borel coupling constant $g$. Now (\ref{mcreg}) has a finite 
radius of convergence and it is possible to analytically continue $\mc^\bb$ to 
the whole complex $g$ plane(usually referred to as Borel plane). To go back to the original 
answer one performs the Laplace transform:
\beq
\label{bint}
\tilde{\mc} = \dfrac{1}{g_0} \int_0^{+\infty} \ dg \ e^{-g/g_0} \mc^\bb(g)
\eeq
If the original series were convergent $\tilde{\mc}$ coincides with $\mc$. However, in most 
interesting physical cases (\ref{bint}) is divergent: $\mc^\bb$ has poles on the positive real axis. 
Ambiguity in the choice of contour leads to the non-perturbative complex ambiguity $\sim 2 \pi i e^{-\text{const}/g_0}$ in
the final answer $\tilde{\mc}$. Nonetheless, we expect(or at least hope) that in sensible theories this ambiguity will be
canceled somehow, probably with some ambiguity in the non-perturbative sector. This is where the BZJ prescription 
enters the game.

It has been known for a very long time that in quantum mechanics the ambiguity coming from 
the leftmost pole is canceled by an ambiguity in the instanton--anti-instanton contribution. This non-perturbative ambiguity arises because instanton and anti-instanton attract
and one can not assume that they are well-separated. Technically the problem is that the integral over the distance 
between the instantons is not well defined. Fortunately, there is a way out: we can analytically 
continue the expression for the instanton--anti-instanton amplitude to the negative 
values of the coupling constant\cite{bm,zj}. Then, instead of attraction we 
will have a repulsion between instanton and anti-instanton and we are able to compute the integral over the
separation. Then, in the final answer, we rotate the coupling constant back to the original positive value. 
This is known as Bogomolny--Zinn-Justin prescription.
However, the final answer is not analytic in $g_0$: typically it has a logarithmic branch 
cut $\sim \log(-g_0) e^{-\text{const}/g_0}$. So rotating $g_0$ back will produce a complex ambiguity which cancels the 
perturbative ambiguity. Since the two ambiguities cancel in the final answer, they should also cancel before we perform
the inverse Borel transform. It means that the instanton--anti-instanton contribution manifests itself on the Borel plane
as a simple pole on the positive real axis. This is exactly what we will see for a particle in double-well potential.

This paper is organized as follows: in section \ref{gen} we will present a general definition of a Borel transformation
on a Lagrangian level and discuss some simple properties of this transformation. Also, we will briefly mention other
interpretations of the Borel transformations apart from coupling to a dilaton. In section \ref{ex} we will apply
our new technique for studying instanton contributions for quantum particle in the double-well potential. We will demonstrate
that in this simple problem our approach reproduces the BZJ prescription. Also we will briefly discuss a field theoretic model
of scalar field with self-interaction. 

\section{Borel transformation}
\label{gen}
We can rewrite eq. (\ref{mcreg}) using the inverse Laplace transform:
\beq
\label{mcint}
\mc^\bb(g) = \cfrac{1}{2 \pi i} \oint_\cc \ \dfrac{dz}{z} e^z \ \sum_{n=0}^{+\infty} \dfrac{c_n g^n}{z^n}
\eeq
where the contour $\cc$ lies to the right of all integrand singularities. Because of the term $e^z$ we can push
the contour to the left, picking up the residue at $z=0$. However, as we will see later, the choice of contour is a very subtle issue. Obviously the perturbative answer ($\ref{mcint}$) corresponds to the following Lagrangian theory:
\beq
\label{zbb}
Z^\bb =\cfrac{1}{2 \pi i} \int_\cc \ \dfrac{dz}{z} \dc x \exp \l i S_0(x) +i S_i \l x,\cfrac{g}{z} \r + z\r
\eeq
This formula is the main result of this paper. As usual with the (inverse) Laplace transform we have to 
understand the integral over $z$ in the following
manner: we compute the integrand in some region in $z$ for which it is well-defined, 
analytically continue it to the complex $z$ plane and carefully pick up a contour to ensure convergence.
The integrand in this case is given by:
\beq
\int \dc x \exp \l i S_0(x) + S_i(x,\cfrac{g}{z}) \r
\eeq
Since the coupling in the new theory enters with the complex factor $1/z$ it can thought of as the
coupling defined in the complex plane.
This justify the natural use of BZJ prescription in the new theory. As we will see the instanton--anti-instanton contribution is well-defined for negative $z$ and has a branch
cut on the positive real axis. The choice of contour is unique in this case.

As we mentioned in the Introduction, this procedure allows to systematically generalize BZJ procedure for the case of several
coupling constants. In this case we will have to introduce a complex field $z_i$ for each coupling constant $\la_i$ and then
study the multi-dimensional complex integral over $z_i$.

Note that if we restore the dependence on the Planck constant the last field 
independent term in the exponent is subleading term. 


Path integral (\ref{mcint}) resembles several issues. First, it looks like the Polyakov action, where $z$ plays the role
of the einbein. 
However, in our case this einbein corresponds not to the time evolution, but rather to the renormalization group(RG) flow. Recently it was suggested that RG flow and resurgence are directly related \cite{whitham,bridge}.
Secondly it reminds the Lipatov action \cite{lipatov}. It also looks like coupling the original theory to the non-dynamical complex dilaton field $\phi=\log(z)$, since we can rewrite the partition function as:
\beq
\label{zbbdil}
Z^\bb = \cfrac{1}{2 \pi i} \int \ d\phi \dc x \exp \l i S_0(x) + i S_i(x,g e^{-\phi})+e^\phi \r
\eeq
The fact that $\phi$ has a canonical dilaton measure and the integration over $\phi$ stands next to the integration over
other field strongly suggests that one has to interpret $\phi$ as an additional field 
or target space coordinate and not an extra coordinate on the worldline or worldvolume.

Let us return to the question of choosing the contour. Assume for a moment that the interaction 
term depends linearly on the coupling constant and let us try to go back
to the original theory by performing direct Laplace transformation. Moreover, let us perform a Wick
rotation and consider the Euclidean action:
\beq
\begin{split}
\int \dc x \dfrac{dz}{z} \dfrac{d g}{g_0} \exp \l  -S_0(x) - \cfrac{g}{z} S_i(x) + z - \dfrac{g}{g_0} \r =
\int \dc x \dfrac{dz}{z + g_0 S_i(x)} \exp \l  - S_0(x) + z \r = \\
2 \pi i \int \dc x \exp \l  - S_0(x) -  g_0 S_i(x) \r 
\end{split}
\eeq
Note that we had to push the contour as far right as $-g_0 S_i(x)$.
This simple formal calculation reveals the following feature: the choice of contour may depend on the field configuration
we are considering. It means that the dilaton field interacts with the original theory not only through the terms in the
Lagrangian, but also through the choice of the contour, that is,  topologically.

The additional integral over $z$ in the partition function implies new relations 
generated by the invariance under the reparametrization of integration variable $z\rightarrow F(z)$\footnote{We are
grateful to A. Gorsky for suggesting this idea.}.
These Ward-like identities are familiar as Virasoro constraints or loop equations
in the matrix models. In quantum field theory language these are known as the Schwinger-Dyson equations.
Consider for example  the simplest transformation $z\rightarrow z(1+\delta \eta)$.
Since the measure is invariant and the answer should not depend on $\delta \eta$, in the linear order in $\delta \eta$
we derive the following relation: 
\beq
\label{loop}
\cfrac{1}{2 \pi i}\int \dc x \cfrac{dz}{z} \l z + z \la \cfrac{\pr S_i(x,\la/z)}{\pr \la}  \r \exp(-S_0(x)-S_i(x,\la/z)+z) =0
\eeq
This equation has a simple meaning in perturbation theory: the perturbative expansion of the partition function is just the 
Taylor expansion of interaction term $\exp(-S_i(x,\la/z))$ in powers of $\la$. Equation (\ref{loop}) means that 
each additional term contains additional inverse factor of $z$. This becomes even more clear if we assume that the interaction
term depends linearly on $\la$. In this case we have:
\beq
\cfrac{1}{2 \pi i}\int \dc x \cfrac{dz}{z} \l z + \la S_i(x) \r \exp(-S_0(x)- \la S_i(x)/z +z) =0
\eeq
It would be interesting to study this relation in the non-perturbative sector or obtain more complex constraints 
arising from more complicated reparametrizations.

 

\section{Instanton contributions}
\label{ex}
In this section we will study how instanton contributions manifest themselves on the Borel plane using the regularized theory.
We will consider two simple models: quantum particle in the double well potential\cite{abc} and massless scalar field theory.
For the quantum mechanical particle the Lagrangian in the Euclidean space reads as:
\beq
\lc = \dfrac{\dot x^2}{2} + \la_0(x^2-\eta^2)^2 =  \dfrac{\dot x^2}{2}-\cfrac{\om^2 x^2}{4}+\la_0 x^4 + \dfrac{\om^4}{64 \la_0}
\eeq
where $\om^2=8 \la_0 \eta^2$. We will keep $\om$ constant and couple dilaton to the interaction constant $\la_0$.

We can study different matrix elements: we can either consider the vacuum energy or the transition amplitude from one
minimum of the potential to the other one. The former case corresponds to the trivial topological sector, 
whereas the later case the topological charge changes by one. The structure of the Borel plane 
will be different, of course. We will find that single instantons manifest themselves as 
branch cuts on the Borel plane, whereas instanton--anti-instanton pairs are related to poles. 
This suggests possible holographic interpretation\footnote{This idea was brought up by A. Gorsky}: if we lift
this quantum mechanical problems to 2 dimensions, instantons will become normal particles. In this case we expect that
single particle excitations are related to branch cuts, whereas instanton--anti-instanton pairs, as bound states,
are related to poles.

Let us first study single instantons.
One instanton solution reads as\cite{abc}:
\beq
x_{\text{inst}}(t)=\eta \tanh \l \dfrac{\om(t-t_c)}{2} \r
\eeq
And the action equals to:
\beq
S_\inst(\la_0) = \dfrac{\om^3}{12 \la_0}
\eeq
Apart from a numerical factor, the contribution from the single zero mode  around the instanton solution yields the factor  
$\sqrt{S_\inst}$. In the transformed theory
we will have $\la/z$ instead of $\la_0$
Therefore, one instanton contribution reads as:
\beq
\cfrac{1}{2 \pi i} \int_\cc \dfrac{dz}{z} \sqrt{S_\inst(\la) z} \exp(-z S_\inst(\la)+z)
\eeq
It is well-defined for positive $z$ and has a branch cut on the negative real axis. Therefore contour $\cc$ goes along the imaginary axis and to the right from the cut. If Borel coupling constant $\la$ is small, so that 
$S_\inst(\la) > 1$ then the endpoints of
the contour $\cc$ have to lie at $z \ra +\infty$. So we can move $\cc$ to the right so that the integral vanishes. In the "strong coupling" region
$S_\inst(\la)<1$, endpoints have to end at $z \ra -\infty$ so we have to integrate over the branch cut. The answer is:
\beq
\dfrac{\sqrt{S_\inst(\la)}}{\sqrt{\pi}\sqrt{(1-S_\inst(\la))}}
\eeq
Therefore in the double-well potential the single instanton manifests itself on the Borel plane as branch cuts.

Let us study now instanton--anti-instanton amplitude. It looks as follows\cite{bm}:
\beq
\label{two}
S_\inst(\la_0) \exp(-2 S_\inst(\la_0)) I(\la_0)
\eeq
where 
\beq
I(\la_0)=\int_0^{+\infty} \ dt \l \exp \l \cfrac{\xi}{\la_0} e^{-\xi t} \r - 1  \r = \int_0^{\xi/\la_0} \ dy (e^y-1)/y
\eeq
The integral over $dt$ is the integral over the separation between instanton and anti-instanton. $\xi$ is a constant 
which does not depend on the coupling constant $\la_0$. 
In the transformed theory instanton--anti-instanton contribution reads as:
\beq
\cfrac{1}{2 \pi i}\int_\cc \cfrac{dz}{z} S_\inst(\la/z) \exp(-2 S_\inst(\la/z)+z) I(\la/z)
\eeq
It is easy to see that the function $I(x)$ has a logarithmic 
branch cut:
\beq
I(x+i0)-I(x-i0) = \oint dy (e^y-1)/y = 2 \pi i
\eeq
The position of the branch cut can be determined by the following consideration.
Formula $(\ref{two})$ was derived assuming 
that the instanton and anti-instanton are well-separated. For positive $\la/z$ the integral
is dominated by small $t$ region, which invalidates the assumption. Therefore we conclude that the cut
should lie on the positive real axis. Note that we did not have
to complexify the coupling constant $\la$ - what we had to do is to pick up a different integration cycle for $z$.

Therefore the contour $\cc$ goes along the imaginary axis and to the left from the cut. Interesting feature of the 
instanton--anti-instanton contribution is that it has a non-vanishing contribution even in the 
"weak coupling region" $2 S_\inst(\la)>1$. Indeed, the only way to ensure convergence in this case is to close the 
contour in $ \Re z > 0$ region and to integrate over the cut. The answer reads as:
\beq
\int_0^{+\infty} \cfrac{dz}{z}  z S_\inst(\la) \exp(-2 z S_\inst(\la)+z) = \cfrac{S_\inst(\la)}{2S_\inst(\la)-1}
\eeq
Similar answer was obtained in \cite{baba} by different means. As we have expected instanton--anti-instanton pair produces a pole on the Borel plane.

As we mentioned before, we expect that after the Borel transformation the non-perturbative sector will be richer. 
Let us illustrate this statement by considering a simple scalar field theory in $D$ Euclidean dimensions:
\beq
\label{scalar}
\lc_E = (\pr b)^2 + \cfrac{\la_0 b^n}{n!}
\eeq
with $D=2n/(n-2)$. We expect that the theory is well-defined for positive $\la_0$. Non-perturbative solutions in this case
are unknown. Now we will show that the Borel regularized theory does have instanton-like solutions. Actually, they coincide
with the so-called Lipatov solution discovered long ago \cite{lipatov} in a bit different setup. 

From the work of Lipatov\cite{lipatov} we know that for $D=4$ the perturbation series in the theory at hand 
are divergent and the
Borel-resummed series contain a simple pole on the negative real axis. 
Lipatov computed the perturbation theory 
asymptotics indirectly: in order to compute $\la_0^N$ term in the 
partition function expansion one can consider the following contour integral:
\beq
\label{lip}
\oint \ d \la_0 \ \mathcal{D} b \cfrac{1}{\la_0^{N+1}} \exp \l - \int d^4 x \mathcal{L}_E \r
\eeq
and then evaluate it using the saddle-point expansion for large $N$. Corresponding solution is known as "Lipaton":
\beq
\label{sol}
b_L = \cfrac{C}{(x-x_0)^2+\rho^2}, \ C,\rho,x_0 \ \text{are constants}
\eeq
The saddle-point value of $\la_0$ is negative. This is the reason why Lipatons exist.


Let us now switch to the Borel transformation. If we apply the Borel regularization for (\ref{scalar}) we will obtain
\beq
\label{my}
\int \dc b \cfrac{dz}{z} \exp \l -\int d^4 x \l (\pr b)^2 + \cfrac{\la b^n}{z n!} \r + z \r
\eeq
which is similar but not equal to (\ref{lip}). Moreover unlike the Lipatov case, we are obliged to integrate over $z$ exactly. 
Nevertheless, now we will show that the Borel regularized theory admits Lipaton solution.
Indeed, this statement is almost trivial: since now we are considering the whole complex $z$-plane we can assume 
that $z$ is negative and then analytically continue along the appropriate contour $\cc$. But for 
any negative values of $z$, $b$ equations of motion
\beq
\pr^2 b = \cfrac{\la b^{n-1}}{n! z}
\eeq
admit Lipatov solution (\ref{sol}). It means that the pole on the negative real axis will be canceled after all. 

\section{Conclusion}
In this paper we defined a Borel transformation of a general Lagrangian theory. 
On the perturbative level it reduces to the Borel regularization, whereas on the non-perturbative level it 
makes the BZJ prescription more natural.
Of course, there are a lot of questions to be answered. For example, we expect that it would be easier 
to see the cancellation of the renormalon singularities in the transformed theory, compared in the original one. Recently it was explicitly shown how to cancel this singularities in a number of quantum field theories\cite{du1,du2}. 
Also it would be quite interesting to make the complex dilaton dynamical and study the properties of the resulting theory.

\section{Acknowledgment}
We would like to thank A. Gorsky for collaboration on the early stages and numerous discussions.
Also we are thankful to K. Bulycheva, V. Kirilin, I. Klebanov, A. Polyakov, G. Tarnopolskiy and P. Wytaszczyk 
for comments and discussions. A.M. is grateful to RFBR grant 15-02-02092 for travel support.

\printbibliography

\end{document}